\documentclass[10pt,letterpaper]{article} 
\usepackage{opex3}
\usepackage{amssymb,amsmath,cite,unit}
\usepackage{bm}

\begin{document}

\title{An invisible medium for circularly polarized electromagnetic waves}

\author{Y.~Tamayama,$^{1,*}$
T.~Nakanishi,$^{1,2}$ K.~Sugiyama,$^{1,2}$ 
and M.~Kitano$^{1,2}$}

\address{$^1$ Department of Electronic Science and 
Engineering, Kyoto University, Kyoto 615-8510, Japan\\
$^2$ CREST, Japan Science and Technology Agency, Tokyo 102-0075, Japan
}

\email{tama@giga.kuee.kyoto-u.ac.jp} 



\begin{abstract}
We study the no reflection condition for
a planar boundary between vacuum and an isotropic chiral medium. 
In general chiral media, elliptically polarized waves incident at a 
particular angle satisfy the no reflection condition. 
When the wave impedance and wavenumber of the chiral medium are equal 
to the corresponding parameters of vacuum, 
one of the circularly polarized waves is transmitted 
to the medium without reflection or refraction for all angles of incidence. 
We propose a circular polarizing beam splitter as a 
simple application of the no reflection effect. 
\end{abstract}




\section{Introduction}

When an electromagnetic (EM) wave is incident on a boundary between two media, the incident wave is partially reflected. 
However, at a particular angle of incidence, 
the reflected wave vanishes. 
This phenomenon is known as the Brewster effect~\cite{saleh07}. 
The Brewster no-reflection 
effect is utilized in many applications; for example, 
intracavity elements are placed at Brewster angles in order to suppress 
insertion losses. 

The Brewster effect for transverse-magnetic (TM) waves (p waves) arises 
at an interface between two media whose permittivities are 
different from each other. 
The TM Brewster effect can be observed in naturally occurring 
dielectric media. 
The Brewster effect for transverse-electric (TE) waves (s waves) 
arises at an interface between two media
whose permeabilities are different from each other~\cite{doyle80,futterman95,fu05,tanaka06,shu07}. 
Normally, it is difficult to observe the Brewster effect for TE 
waves in naturally occurring 
media because they do not respond to magnetic 
fields in high frequency regions, i.e., 
microwave, terahertz, and optical regions. 
However, magnetic media can be fabricated in high frequency regions by using 
metamaterials~\cite{pendry99, marques03, enkrich05, grigorenko05, xu06}, 
and therefore, the TE Brewster effect can be observed. 
The effect has been experimentally confirmed 
in the microwave region~\cite{tamayama06} and 
also in the optical region~\cite{ishihara08}. 

In addition to permittivity and permeability, it is possible 
to control the chirality 
parameter and the non-reciprocity parameter 
by using metamaterials; therefore, it is important to explore the 
no reflection conditions for generalized media. 
Brewster conditions for various media have been
studied by several researchers~\cite{doyle80, azzam83, kim86, bassiri88, lakhtakia89, lakhtakia92, futterman95, leskova03, fu05, grzegorczyk05, tanaka06, tamayama06, shu07, ishihara08}. 
Some researchers have derived the expressions of 
Brewster conditions for chiral and non-reciprocal media. 
For these media, the incident polarization for no reflection is not 
necessarily perpendicular nor parallel to the plane of incidence. 
The no reflection conditions cannot be found if the incident wave 
is assumed to be TM or TE waves. 
It has been pointed out that the Brewster condition must be revised such 
that the polarization of the reflected wave is independent of the 
incident polarization~\cite{bassiri88, lakhtakia89, lakhtakia92}. 
However, thus far, the explicit relations among the medium parameters for 
achieving non reflectivity have not been determined. 

The objective of the present study is to explicitly derive 
the no reflection conditions 
for an (isotropic) chiral medium that responds to the electric and 
magnetic fields simultaneously. 
This class of medium, having 
relative permittivity $\varepsilon \sub{r} \neq 1$,  relative permeability $\mu \sub{r} \neq 1$, and normalized chirality 
parameter 
$\xi \sub{r} \neq 0$, can be accessible with the current 
technology of metamaterials. 

First, we confirm that the revised Brewster condition~\cite{bassiri88, lakhtakia89, lakhtakia92} reduces to the condition that the reflection 
(Jones) matrix has at least one vanishing eigenvalue. 
We show that the analysis can be largely simplified by using 
Pauli matrices~\cite{sakurai94}. 

It is found that in general chiral media, the no reflection condition is 
satisfied by elliptically polarized incident waves having a particular 
angle of incidence. 
This is merely a natural extension of the  usual Brewster effect for 
achiral ($\xi \sub{r} =0$) media. 
In addition, a qualitatively new mode of no reflection is found. 
When the wave impedance and wavenumber 
of the chiral medium, determined by 
$\varepsilon \sub{r} ,~ \mu \sub{r}$, and $\xi \sub{r}$, are 
equal to the vacuum values $Z_0$ and $k_0$, respectively, 
one of the circularly polarized waves is transmitted to 
the medium without either reflection or refraction. 
The no reflection condition is independent of the incident angle, 
i.e., the medium is totally transparent with respect to only that 
circular polarization. 
The other circularly polarized (CP) wave is refracted and reflected, or 
even totally reflected. 
The no reflection 
phenomenon can be physically understood as a destructive interference 
of electric and magnetic responses, due to the mixing through the chirality 
parameter. 

A simple, straightforward application of the totally transparent medium 
for the circularly polarized wave is a circular polarizing beam splitter 
(CPBS). 
We analyze the CPBS by a finite-difference time-domain (FDTD) method~\cite{taflove05}. 
The properties of this CPBS are almost ideal compared with 
earlier polarizing beam splitters~\cite{jacobs88, mahmoud98, davis01, azzam03}.

\begin{figure}[tb]
\begin{center}
\includegraphics{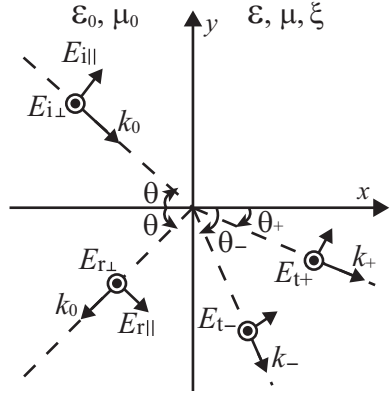}
\caption{Geometry of coordinate system. The incident, reflected, and transmitted waves 
are denoted by the subscripts i, r, and t, respectively. 
Region $x<0$ represents vacuum, and region $x \geq 0$ represents 
the chiral medium. }
\label{fig:coordinate}
\end{center}
\end{figure}

\section{No reflection condition for chiral media}

As shown in Fig.~\ref{fig:coordinate}, we suppose that an EM wave 
with an angular frequency $\omega$ 
is incident from vacuum (permittivity $\varepsilon _0$, 
permeability $\mu _0$) on an isotropic chiral medium 
at an incident angle of $\theta$. 
The constitutive equations for the chiral medium are given as~\cite{lakhtakia98,serdyukov01,monzon05,qiu08}
\begin{equation}
\bm{D} 
= 
\varepsilon \bm{E} - \ii \xi \bm{B} ,\quad
\bm{H}
=
\mu ^{-1} \bm{B} - \ii \xi \bm{E} , \label{eq:1-10} 
\end{equation}
where $\varepsilon$, $\mu$, and $\xi$ are the permittivity, 
permeability, and chirality parameter, respectively. 
We adopt the EB formulation for the constitutive equations, but 
it is also possible to rewrite them in terms of EH
formulation~\cite{lakhtakia92}. 
Due to the translational invariance of the interface, 
Snell's equations $k_0 \sin{\theta} = k_+ 
\sin{\theta _+} = k_- \sin {\theta _-}$ are satisfied. 
Here, $k_0 = \omega \sqrt{\varepsilon _0 \mu _0}$ 
is the wavenumber in vacuum, $ k_{\pm} = \omega 
( \sqrt{\varepsilon \mu + \mu ^2 \xi ^2} \pm \mu \xi )$ are 
the wavenumbers for left circularly polarized (LCP) and 
right circularly polarized (RCP) waves in the chiral medium, and 
$\theta _+ ~ (\theta _- )$ is the refractive angle of the LCP (RCP) 
wave. 

The relation between the electric field of the incident wave, 
$\bm{E} \sub{i} = [E_{\mathrm{i}\perp} , E_{\mathrm{i}\parallel} ]^T$ 
($T$ stands for transposition), and that of the reflected wave, 
$\bm{E} \sub{r} = [E_{\mathrm{r}\perp} , E_{\mathrm{r}\parallel} ]^T$, 
is written as~\cite{bassiri88}
\begin{align}
\bm{E} \sub{r}
&=
\frac{1}{\varDelta}
M\sub{R}
\bm{E} \sub{i}, \quad
M\sub{R}
=
c\sub{u} I + c_2 \sigma _2 + c_3 \sigma _3 ,
\label{eq:1-30} \\
c\sub{u}
&=
2 Z_0 Z\sub{c} ( \cos^2{\theta} - \cos{\theta _+} \cos{\theta _-} ) ,
\nonumber \\
c_2
&=
- 2 Z_0 Z\sub{c} \cos{\theta} ( \cos{\theta _+} - \cos{\theta _-} ) ,
\nonumber \\
c_3
&=
( Z\sub{c} ^2 - Z_0 ^2) \cos{\theta} ( \cos{\theta _+} + \cos{\theta _-} ), 
\nonumber \\
\varDelta
&=
 ( Z\sub{c} ^2 + Z_0 ^2 ) \cos{\theta} ( \cos{\theta _+} + \cos{\theta _- } ) 
 + 2 Z_0 Z\sub{c} ( \cos ^2{\theta} + \cos{\theta _+} \cos{\theta _-} ) ,
\nonumber
\end{align}
where 
$
Z_0
=
\sqrt{\mu _0 / \varepsilon _0}
$ 
and 
$
Z\sub{c}
=
\sqrt{ \mu / ( \varepsilon + \mu \xi ^2 )}
$
are the wave impedances of vacuum and the chiral medium, respectively. 
We introduce the unit matrix $I$ and the Pauli 
matrices~\cite{sakurai94}: 
\begin{equation}
\sigma _2
=
\begin{bmatrix}
0 & -\ii \\
\ii & 0
\end{bmatrix}, \quad
\sigma _3
=
\begin{bmatrix}
1 & 0 \\
0 & -1
\end{bmatrix}.
\end{equation}
The reflection matrix $M\sub{R}$ can be rewritten as 
\begin{equation}
M\sub{R}
=
c\sub{u} I + c_{\varphi} \sigma _{\varphi}, \label{eq:5-10}
\end{equation}
where
$
c_{\varphi}
=
\sqrt{ c_2 ^2 + c_3 ^2},~
\sigma _{\varphi}
=
\sigma _2 \sin{\varphi}  +  \sigma _3 \cos{\varphi}
,~
\sin{\varphi}
=
c_2 / c_{\varphi}$, and $
\cos{\varphi}
=
c_3 / c_{\varphi}
$. 

The no reflection condition is satisfied when $M\sub{R}$ has a zero 
eigenvalue, namely, $\mathrm{det}\,(M\sub{R}) =0$ or 
$\mathrm{rank}\,(M\sub{R}) \leq 1$. For the 
incident wave with the corresponding eigenpolarization, 
the reflection is nullified. 
From Eq.~(\ref{eq:5-10}), we observe that the eigenvalue problem for 
$M\sub{R}$ reduces to that for $\sigma _{\varphi}$. 
The eigenvalues of $\sigma _{\varphi}$ are $\pm 1$ and their 
corresponding eigenpolarizations 
are $\bm{e} _{\varphi +} = 
\cos{(\varphi/2)} \,\bm{e}_{z} + \ii \sin{(\varphi /2)} 
( \bm{e} _{x} \sin{\theta} + \bm{e} _{y} \cos{\theta} )$ 
and $\bm{e} _{\varphi -} = 
\sin{(\varphi/2)} \,\bm{e}_{z} - 
\ii \cos{(\varphi /2)} 
( \bm{e} _{x} \sin{\theta} + \bm{e} _{y} \cos{\theta} )$, 
where $\bm{e}_x ,~ \bm{e} _y$, and $\bm{e} _z$ are the unit vectors 
in the direction of the positive $x$-, $y$-, and $z$-axes, 
respectively. 
Therefore, $M\sub{R}$ has a zero eigenvalue when $c\sub{u} = c_{\varphi}$ 
($c\sub{u} = - c_{\varphi}$)
is satisfied, and 
no reflection is achieved for the incident wave with the 
polarization 
$\bm{e} _{\varphi -}$ ($\bm{e} _{\varphi +}$).

{\it Achiral case} ($\xi =0$)---The reflection matrix is 
$M\sub{R} = c\sub{u} I + c_3 \sigma _3$. 
The eigenpolarizations are $\bm{e}_{x} \sin{\theta} + \bm{e}_y 
\cos{\theta}$ and $\bm{e}_{z}$; 
therefore, the no reflection 
condition can be satisfied only 
for linearly polarized (LP) waves. 
The no reflection effect is observed at a particular incident angle 
$\theta$ that satisfies 
$c\sub{u} = \pm c_3$. 
The condition $c\sub{u} = c_3$ ($c\sub{u} = - c_3$) yields 
a no-reflection angle, 
called the Brewster angle, for TM (TE) waves in isotropic achiral media. 

{\it Chiral case} ($\xi \neq 0$)---First, we consider a 
general case of $Z\sub{c} \neq 
Z_0$, which gives $\varphi \neq n\pi /2$ with an integer $n$. The eigenpolarizations are $\bm{e} _{\varphi \pm}$; hence, the no reflection 
condition can be satisfied only for elliptically polarized (EP) waves. 
The no reflection effect is observed 
at a particular incident angle $\theta$ 
satisfying $c\sub{u} = \pm c_{\varphi}$, 
which is a natural extension of the usually observed no reflection effect, 
or the Brewster effect in achiral media. 

When $Z\sub{c} = Z_0$, 
we have $M\sub{R} = c\sub{u} I + c_2 \sigma _2$. 
The eigenpolarizations are 
$[\bm{e}_{z} \pm \ii  ( \bm{e}_{x} \sin{\theta} + \bm{e}_y 
\cos{\theta} ) ] / \sqrt{2}$; 
hence, the no reflection 
condition can be satisfied only for CP waves. 
The condition $\theta _+ = \theta $ ($\theta _- = \theta $) is 
required in order 
to satisfy $c\sub{u} = -c_2$ ($c\sub{u} = c_2$), 
which is the no reflection condition for 
LCP (RCP) waves. 
We note that once $k_+ = k_0$ ($k_- = k_0$) is satisfied by selecting the 
constants of medium, 
$c\sub{u} = -c_2$ ($c\sub{u} = c_2$) is satisfied for any $\theta$. 
Namely, 
the no reflection condition is satisfied for all angles of incidence. 
This observation is quite different 
from the no reflection conditions for TM and TE 
waves in isotropic achiral media and for EP waves in isotropic chiral 
media. 
A qualitatively new mode of no reflection is obtained for LCP (RCP) waves
in isotropic chiral media when 
the wave impedance matching condition $Z\sub{c} = Z_0$ and the wavenumber 
matching condition $k_+ = k_0$ ($k_- = k_0$) are satisfied simultaneously. 

\begin{figure}[tb]
\begin{center}
\includegraphics{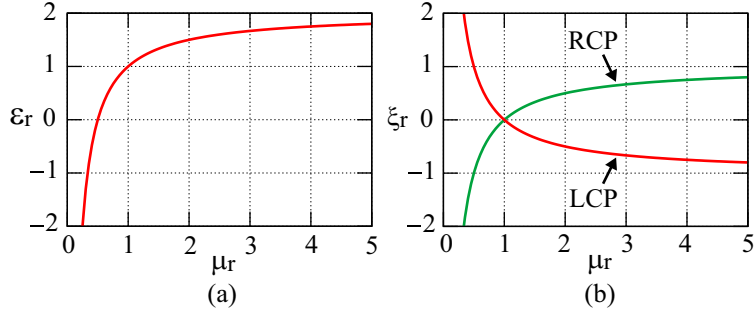}
\caption{(a) Relation between $\mu \sub{r}$ and $\varepsilon \sub{r}$ and 
(b) between $\mu \sub{r}$ and $\xi \sub{r}$ for 
no reflection conditions (invisible conditions). 
In the ($\mu \sub{r}$-$\xi \sub{r}$) graph, 
the red and green lines represent the conditions for LCP and RCP
waves, respectively. }
\label{fig:noref}
\end{center}
\end{figure}

We derive the explicit relations among $\varepsilon ,~ \mu$, and $\xi$ for 
the no reflection condition for CP waves. 
From the above discussion,  
both $Z\sub{c} = Z_0$ and $k_+ = k_0$ ($k_- = k_0$) 
are necessary 
and yield the following relations: 
\begin{equation}
\varepsilon \sub{r}
=
2 - \frac{1}{\mu \sub{r}} ,\quad 
\xi \sub{r}
=
\mp \left( 1 - \frac{1}{\mu \sub{r}} \right)  ,
\label{eq:2-150} 
\end{equation}
where
$\varepsilon \sub{r} = \varepsilon / \varepsilon _0 ,~
\mu \sub{r} = \mu / \mu _0 $ and $
\xi \sub{r} = Z_0 \xi $. 
The negative (positive) sign in Eq.~(\ref{eq:2-150}) indicates 
the condition for 
LCP (RCP) waves. 
Figure \ref{fig:noref}(a) represents the ($\mu \sub{r}$-$\varepsilon 
\sub{r}$) relation shown in Eq.~(\ref{eq:2-150}), and 
Fig.~\ref{fig:noref}(b) shows the ($\mu \sub{r}$-$\xi \sub{r}$) 
relation. 
It should be noted that the no reflection condition can be satisfied by 
shifting the parameters $(\varepsilon \sub{r}, \mu \sub{r}, \xi \sub{r})$ 
away from the vacuum parameters $(1,1,0)$ by a small amount. 
Such a medium can be obtained by using the state-of-art technology of 
metamaterials. 

It is already known that under certain conditions, 
the no reflection effect for LP waves is observed at any incident angle 
in {\it anisotropic} achiral media~\cite{shu07}. 
In this case, 
the incident wave is refracted in the anisotropic medium, 
while for the present case, 
the incident wave is transmitted straight through the chiral medium. 
Thus, the medium with $Z\sub{c} = Z_0$ and $k_+ = k_0$ ($k_- = k_0$) can be 
considered as vacuum, namely, 
the medium is invisible, for LCP (RCP) waves.

\section{Application of invisible medium for CP waves}

\begin{figure}[tb]
\begin{center}
\includegraphics{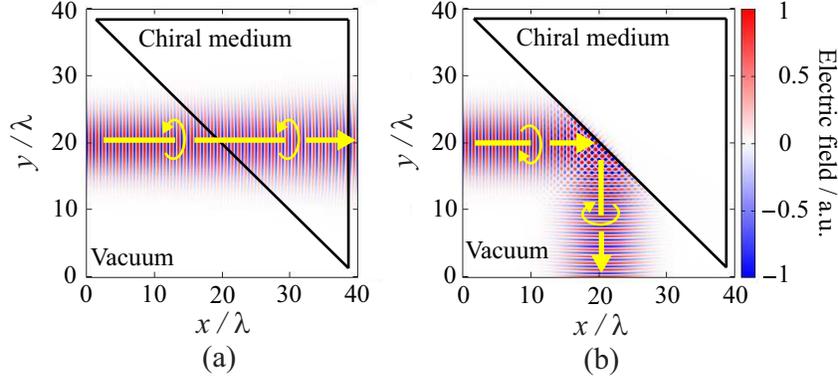}
\caption{Results of two-dimensional FDTD analysis of the CPBS. 
Propagation of (a) LCP waves and (b) RCP waves. 
Straight and curved arrows represent the propagation direction and 
polarization direction, respectively. 
$\lambda$ is the wavelength of the EM waves. }
\label{fig:splitter}
\end{center}
\end{figure}

We propose a CPBS 
as one of the applications of the invisible medium for CP 
waves. 
Here, we set the parameters of the chiral medium as 
$\varepsilon \sub{r} = 0.75,~\mu \sub{r} = 0.8$, and $\xi \sub{r} = 0.25$, 
which give the invisible condition for LCP waves. 
For $k_- = 0.6 k_0$, Snell's equation for RCP waves is expressed as 
$\sin{\theta} = 0.6 \,\sin{\theta _-}$; hence, 
the critical angle for RCP waves is $\theta \sub{c} = \arcsin{(0.6)} 
\simeq 37^{\circ}$. 
Therefore, LCP waves are completely transmitted without any reflection, 
while RCP waves
are totally reflected with the incident angle greater than $37^{\circ}$. 
This implies that we can divide the incident waves into LCP and RCP waves. 

We carry out an FDTD analysis 
\cite{taflove05} of the CPBS\@. 
It is assumed that the EM wave transmitted from vacuum 
is incident at an angle of $45^{\circ}$ ($> \theta \sub{c}$) 
on a triangular prism made of the chiral medium. 
In order to adopt the two-dimensional FDTD method, Maxwell's equations 
for CP waves are rearranged as follows: 
\begin{align}
\fracpd{E_z}{y} 
&= 
\ii \omega ( \mu \pm \mu \xi Z\sub{c} ) H_x , \\
-\fracpd{E_z}{x}
&=
\ii \omega ( \mu \pm \mu \xi Z\sub{c} ) H_y , \\
\fracpd{H_y}{x} - \fracpd{H_x}{y}
&=
-\ii \omega \left[ (\varepsilon + \mu \xi ^2 ) 
\pm \frac{\mu \xi}{Z\sub{c}}
  \right] E_z ,
\end{align}
where the relation $\bm{H} = \pm (\ii \, / Z\sub{c}) 
\bm{E}$~\cite{bassiri88} is used; and the positive (negative) sign 
corresponds to LCP (RCP) waves. 
Figure \ref{fig:splitter} shows the results 
(a) for LCP waves and (b) for RCP waves. 
It is observed that the LCP wave is transmitted straight through the chiral 
medium without any reflection and that 
the RCP wave is totally reflected at the surface of the chiral medium. 
It is confirmed that the incident wave can be split 
into LCP and RCP waves and circular polarizing beam splitter is achieved. 

The advantages of the CPBS are as follows. 
A wide acceptance angle is obtained ($53^{\circ}$ in the above mentioned
example). 
An incident wave with arbitrary polarization is distinctly split into 
LCP and RCP wave components with no losses. 
Anti-reflection coating is not required. 
A single element, i.e., one chiral prism, is sufficient for the CPBS. 
Although frequency sensitivity depends on material dispersion, 
broadband metamaterials make the realization of a 
frequency insensitive CPBS possible. 
The ideal properties of the CPBS make it an efficient 
polarizing beam splitter.

\section{Physical meaning of invisible condition for CP waves}

We consider the physical meaning of 
the invisible condition for CP waves in the chiral medium. 
For simplicity, let us assume that 
the invisible condition is satisfied for 
LCP waves. 

First, we consider the medium 
polarization $\bm{P}$ and magnetization $\bm{M}$ induced 
by $\bm{E}$ and $\bm{B}$ in LCP waves. 
They are given by
$
\bm{P}
=
\bm{P} \sub{E} + \bm{P} \sub{B}$
and
$
\bm{M}
=
\bm{M} \sub{B} + \bm{M} \sub{E} 
$,
where $\bm{P} \sub{E} = ( \varepsilon - \varepsilon _0 ) 
\bm{E}$, $\bm{P} \sub{B} = - \ii \, \xi \bm{B}$, $\bm{M} \sub{B} 
= - ( \mu ^{-1} - \mu _0  ^{-1} ) \bm{B}$, and $\bm{M} \sub{E} = 
\ii \, \xi \bm{E}$~\cite{serdyukov01}. 
Taking into account the relation $\bm{H} = ( \ii \, / Z\sub{c}) \bm{E}$ 
that is satisfied for LCP waves~\cite{bassiri88},  
from Eqs.~(\ref{eq:1-10}) and (\ref{eq:2-150}), 
it is not 
difficult to confirm that 
$\bm{P} = 0$ and $\bm{M} =0$ are 
satisfied irrespective of the propagation direction. 
Owing to the electromagnetic 
mixing attributed to $\xi$, 
the polarization $\bm{P} \sub{B}$ 
induced by the magnetic flux density 
completely cancels out 
the polarization $\bm{P} \sub{E}$ induced by the electric field. 
Similarly, $\bm{M} \sub{E}$ cancels out $\bm{M} \sub{B}$. 
As a result of the destructive 
interference of electric and magnetic responses, 
net polarization and magnetization vanish in the case of 
LCP waves in the chiral medium; namely, 
the chiral medium is identical to vacuum for LCP waves. 
Therefore, for any angle of incidence, 
LCP waves are transmitted without reflection or refraction.

\section{Summary and discussion}

We studied the no reflection condition for chiral media. 
In addition to the no reflection effect for EP waves, which is a 
natural extension of the usual Brewster effect for LP waves, 
we found a qualitatively 
new mode of the no reflection effect. 
When the 
wave impedance matching condition $Z\sub{c} = Z_0$ and the wavenumber 
matching condition $k_+ = k_0$ ($k_- = k_0$) were satisfied simultaneously, 
LCP (RCP) waves were transmitted from vacuum to an 
isotropic chiral medium without any reflection or 
refraction irrespective of the incident angle. 
The chiral medium was invisible for LCP (RCP) waves. 

In this paper, we ignore losses of the media for simplicity.
According to our estimation, the loss of metamaterials available today 
is small enough for the demonstration of no reflection phenomena.
In fact, we have examined the TE Brewster condition using 
split ring resonators (SRRs) 
and obtained substantial reduction of reflection, $-27\,\U{dB}$, 
successfully at near resonance frequency~\cite{tamayama06}.
Metamaterials for chiral media can be designed by slightly
modifying the structure of the SRRs~\cite{tretyakov96},
so it will be possible to experimentally demonstrate the no reflection 
effect for circularly polarized wave.

We proposed a CPBS as a straightforward application of the invisible 
medium for CP waves. 
The CPBS was a simple structure, i.e., 
one prism made of the chiral medium, 
and the ideal properties of CPBS make it an efficient 
polarizing beam splitter. 
We believe that the invisible medium for CP waves 
can be used in many other applications. For example, 
we could fabricate invisible containers for CP waves. 
The container can physically confine fluids or gases; however, it is 
invisible for CP waves. 
Another example is 
circular-polarization-selective waveguides that transmit 
only one of the CP waves for which the invisible condition is 
{\it not} satisfied. 

For future studies, it is necessary to prepare metamaterials whose 
$\varepsilon \sub{r} ,~\mu \sub{r}$, and $\xi \sub{r}$ satisfy 
the invisible condition for CP waves. 
Such metamaterials can be realized by employing chiral 
structures~\cite{tretyakov96} in the microwave region and by 
electromagnetically induced chirality in atomic 
systems~\cite{sautenkov05,jurgen07} in the optical region.

\section*{Acknowledgments}

We gratefully thank Vladan Vuleti\'{c} for helpful discussion. 
This research was supported by the Global COE program 
\textquotedblleft Photonics and
Electronics 
Science and Engineering" at Kyoto University.


\begin{thebibliography}{10}
\newcommand{\enquote}[1]{``#1''}
\expandafter\ifx\csname url\endcsname\relax
  \def\url#1{\texttt{#1}}\fi
\expandafter\ifx\csname urlprefix\endcsname\relax\def\urlprefix{URL }\fi
\providecommand{\eprint}[2][]{\url{#2}}

\bibitem{saleh07}
B.~E.~A. Saleh and M.~C. Teich, \emph{Fundamentals of Photonics}, 2nd ed.
  (Wiley-Interscience, 2007).

\bibitem{doyle80}
W.~T. Doyle, \enquote{Graphical approach to Fresnel's equations for reflection
  and refraction of light,} Am.~J.~Phys. \textbf{48}, 643-647 (1980).

\bibitem{futterman95}
J.~Futterman, \enquote{Magnetic Brewster angle,} Am.~J.~Phys. \textbf{63},
  471 (1995).

\bibitem{fu05}
C.~Fu, Z.~M. Zhang, and P.~N. First, \enquote{Brewster angle with a
  negative-index material,} Appl.~Opt. \textbf{44}, 3716-3724 (2005).

\bibitem{tanaka06}
T.~Tanaka, A.~Ishikawa, and S.~Kawata, \enquote{Unattenuated light transmission
  through the interface between two materials with different indices of
  refraction using magnetic metamaterials,} Phys.~Rev.~B \textbf{73},
  125423 (2006).

\bibitem{shu07}
W.~Shu, Z.~Ren, H.~Luo, and F.~Li, \enquote{Brewster angle for anisotropic
  materials from the extinction theorem,} Appl.~Phys.~A \textbf{87},
  297-303 (2007).

\bibitem{pendry99}
J.~B. Pendry, A.~J. Holden, D.~J. Robbins, and W.~J. Stewart,
  \enquote{Magnetism from Conductors and Enhanced Nonlinear Phenomena,} IEEE
  Trans.~Microwave Theory Tech. \textbf{47}, 2075-2084 (1999).

\bibitem{marques03}
R.~Marqu\'{e}s, F.~Mesa, J.~Martel, and F.~Medina, \enquote{Comparative
  Analysis of Edge- and Broadside-Coupled Split Ring Resonators for Metamaterial
  Design---Theory and Experiments,} IEEE Trans.~Antennas Propag.
  \textbf{51}, 2572-2581 (2003).

\bibitem{enkrich05}
C.~Enkrich, M.~Wegener, S.~Linden, S.~Burger, L.~Zschiedrich, F.~Schmidt, J.~F.
  Zhou, Th.~Koschny, and C.~M. Soukoulis, \enquote{Magnetic Metamaterials at
  Telecommunication and Visible Frequencies,} Phys.~Rev.~Lett. \textbf{95},
  203901 (2005).

\bibitem{grigorenko05}
A.~N. Grigorenko, A.~K. Geim, H.~F. Gleeson, Y.~Zhang, A.~A. Firsov, I.~Y.
  Khrushchev, and J.~Petrovic, \enquote{Nanofabricated media with negative
  permeability at visible frequencies,} Nature \textbf{438}, 335-338
  (2005).

\bibitem{xu06}
X.-L. Xu, B.-G. Quan, C.-Z. Gu, and L.~Wang, \enquote{Bianisotropic response of
  microfabricated metamaterials in the terahertz region,} J.~Opt.~Soc.~Am.~B
  \textbf{23}, 1174-1180 (2006).

\bibitem{tamayama06}
Y.~Tamayama, T.~Nakanishi, K.~Sugiyama, and M.~Kitano, \enquote{Observation of
  Brewster's effect for transverse-electric electromagnetic waves in
  metamaterials: Experiment and theory,} Phys.~Rev.~B \textbf{73}, 193104
  (2006).

\bibitem{ishihara08}
R.~Watanabe, M.~Iwanaga, and T.~Ishihara, \enquote{s-polarization
	Brewster's angle of stratified metal-dielectric metamaterial in
	optical regime,} Phys.~Stat.~Sol.~(b) \textbf{245}, 2696-2701
	(2008). 

\bibitem{azzam83}
R.~M.~A. Azzam, \enquote{Maximum minimum reflectance of parallel-polarized
  light at interfaces between transparent and absorbing media,}
  J.~Opt.~Soc.~Am. \textbf{73}, 959-962 (1983).

\bibitem{kim86}
S.~Y. Kim and K.~Vedam, \enquote{Analytic solution of the pseudo-Brewster
  angle,} J.~Opt.~Soc.~Am.~A \textbf{3}, 1772-1773 (1986).

\bibitem{bassiri88}
S.~Bassiri, C.~H. Papas, and N.~Engheta, \enquote{Electromagnetic wave
  propagation through a dielectric-chiral interface and through a chiral slab,}
  J.~Opt.~Soc.~Am.~A \textbf{5}, 1450-1459 (1988).

\bibitem{lakhtakia89}
A.~Lakhtakia, \enquote{Would Brewster recognize today's Brewster angle?}
  Opt.~News \textbf{15}, 14-18 (1989).

\bibitem{lakhtakia92}
A.~Lakhtakia, \enquote{General schema for the Brewster conditions,} Optik
  (Stuttgart) \textbf{90}, 184-186 (1992).

\bibitem{leskova03}
T.~A. Leskova, A.~A. Maradudin, and I.~Simonsen, \enquote{Coherent Scattering
  of an Electromagnetic Wave From, and its Transmission Through, a Slab of a
  Left-Handed Medium with a Randomly Rough Illuminated Surface,} Proc. 
SPIE \textbf{5189}, 22-35 (2003).

\bibitem{grzegorczyk05}
T.~M. Grzegorczyk, Z.~M. Thomas, and J.~A. Kong, \enquote{Inversion of critical
  angle and Brewster angle in anisotropic left-handed metamaterials,}
  Appl.~Phys.~Lett. \textbf{86}, 251909 (2005).

\bibitem{sakurai94}
J.~J. Sakurai, \emph{Modern Quantum Mechanics}, revised ed. (Addison-Wesley,  1994).

\bibitem{taflove05}
A.~Taflove and S.~C. Hagness, \emph{Computational Electrodynamics: The
  Finite-Difference Time-Domain Method}, 3rd ed. (Artech House, 2005).

\bibitem{jacobs88}
S.~D. Jacobs, K.~A. Cerqua, K.~L. Marshall, A.~Schmid, M.~J. Guardalben, and
  K.~J. Skerrett, \enquote{Liquid-crystal laser optics: design, fabrication,
  and performance,} J.~Opt.~Soc.~Am.~B \textbf{5}, 1962-1979 (1988).

\bibitem{mahmoud98}
S.~F. Mahmoud and S.~Tariq, \enquote{Gaussian beam splitting by a chiral
  prism,} J.~Electromagnet.~Wave. \textbf{12}, 73-83 (1998).

\bibitem{davis01}
J.~A. Davis, J.~Adachi, C.~R. Fern\'{a}ndez-Pousa, and I.~Moreno,
  \enquote{Polarization beam splitters using polarization diffraction gratings,}
  Opt.~Lett. \textbf{26}, 587-589 (2001).

\bibitem{azzam03}
R.~M.~A. Azzam and A.~De, \enquote{Circular polarization beam splitter that
  uses frustrated total internal reflection by an embedded symmetric achiral
  multilayer coating,} Opt.~Lett. \textbf{28}, 355-357 (2003).

\bibitem{lakhtakia98}
A.~Lakhtakia, \enquote{Cross-refractive chiral media and constitutive
	contrasts,} Microwave Opt.~Technol.~Lett. \textbf{20}, 337-339
	(1999).  

\bibitem{serdyukov01}
A.~Serdyukov, I.~Semchenco, S.~Tretyakov, and A.~Sihvola,
  \emph{Electromagnetics of Bi-anisotropic Materials: Theory and Applications}
  (Gordon and Breach Science Publishers, 2001).

\bibitem{monzon05}
C.~Monzon and D.~W.~Forester, 
\enquote{Negative Refraction and Focusing of Circularly Polarized Waves
	in Optically Active Media,} Phys.~Rev.~Lett. \textbf{95}, 123904
	(2005). 

\bibitem{qiu08}
C.-W.~Qiu, N.~Burokur, S.~Zouhd, and L.-W.~Li, 
\enquote{Chiral nihility effects on energy flow in chiral materials,}
	J.~Opt.~Soc.~Am.~A \textbf{25}, 55-63 (2008).

\bibitem{tretyakov96}
S.~A. Tretyakov, F.~Mariotte, C.~R. Simovski, T.~G. Kharina, and J.-P. Heliot,
  \enquote{Analytical Antenna Model for Chiral Scatterers: Comparison with
  Numerical and Experimental Data,} IEEE Trans.~Antennas Propag.
  \textbf{44}, 1006-1014 (1996).

\bibitem{sautenkov05}
V.~A.~Sautenkov, Y.~V.~Rostovtsev, H.~Chen, P.~Hsu, G.~S.~Agarwal, and M.~O.~Scully, \enquote{Electromagnetically Induced Magnetochiral Anisotropy in a Resonant Medium} Phys.~Rev.~Lett. \textbf{94}, 233601 (2005). 

\bibitem{jurgen07}
J.~K$\ddot{\mathrm{a}}$stel, M.~Fleischhauer, S.~F. Yelin, and R.~L. Walsworth,
  \enquote{Tunable Negative Refraction without Absorption via
  Electromagnetically Induced Chirality,} Phys.~Rev.~Lett. \textbf{99},
  073602 (2007).

\end{thebibliography}
\end{document}